\documentclass[]{elsart}
\usepackage{psfig}
\usepackage{amssymb}
\def\astrobj#1{#1}

\begin{document}
 \begin{frontmatter}
  \title{On the correlation between the recent star 
         formation rate in the Solar Neighbourhood
         and the glaciation period record on Earth}
  \author{R. de la Fuente Marcos}
   \address{Suffolk University Madrid Campus,
            C/ Vi\~na, 3. E-28003 Madrid, Spain \\
            raul@galaxy.suffolk.es}
  \author{C. de la Fuente Marcos}
   \address{Universidad Complutense de Madrid. E-28040 Madrid, Spain \\ 
            nbplanet@fis.ucm.es}

  \begin{keyword}
         Galaxy: disk - Galaxy: evolution -
         Galaxy: open clusters and associations: general -
         solar neighbourhood - star formation
  \end{keyword}

  \begin{abstract}
         Shaviv (2003) has shown evidence for a correlation between 
         variations in the Galactic cosmic ray flux reaching Earth 
         and the glaciation period record on Earth during the last 
         2 Gyr. If the flux of cosmic rays is mainly the result of 
         Type II supernovae, an additional correlation between the 
         star formation history of the Solar Neighbourhood and the 
         timing of past ice ages is expected. Higher star formation 
         rate implies increased cosmic ray flux and this may 
         translate into colder climate through a rise in the average 
         low altitude cloud cover. Here we reanalyze the correlation 
         between this star formation history and the glaciation 
         period record on Earth using a volume limited open cluster 
         sample. Numerical modeling and recent observational data 
         indicate that the correlation is rather strong but only if 
         open clusters within 1.5 kpc from the Sun are considered. 
  \end{abstract}

 \end{frontmatter}

 \section{Introduction} \label{intro}
    During the past 1 Gyr, the \astrobj{Earth}'s climate has fluctuated
    between warm periods and cold periods or glaciations. The most recent 
    period of glaciation was at its maximum about 10 kyr ago. Although the 
    exact causes of the glacial cycles are not clear, a number of mechanisms 
    able to trigger the cycles of the ice ages have been suggested, ranging 
    from variations in the Solar Cycle (Rind, 2002; Marsh and Svensmark, 
    2003; Scherer and Fichtner, 2004) or in the orbit of the \astrobj{Earth} 
    (Williams et al., 1998) to natural oscillations of the 
    ocean/continent/atmosphere system (Donnadieu et al., 2004). 

    Recently, Shaviv (2001, 2003a) and Svensmark (2004) have put forward 
    evidence for a correlation between variations in the flux of cosmic 
    rays reaching the \astrobj{Earth} and the timing of past ice ages. 
    Cosmic-ray particles, or cosmic rays for short, continuously collide 
    with \astrobj{Earth}, and they make up the only samples of matter from 
    outside the \astrobj{Solar System}. They are sub-atomic particles, 
    nearly 90\% are protons and the rest are nuclei of heavier atoms (8\%) 
    and electrons (1\%). Cosmic rays are very energetic and virtually all 
    of them travel very close to the speed of light. When a cosmic-ray 
    particle hits \astrobj{Earth}'s atmosphere, it creates a series of 
    cascades of many lower-energy particles. These secondary, lower-energy 
    particles in turn collide with objects on \astrobj{Earth}'s surface 
    virtually all the time. Cosmic rays are produced as a result of stellar 
    activity, violent events at the \astrobj{Galactic Center} and at great 
    distances in the hearts of galaxies far beyond the \astrobj{Milky Way}, 
    neutron stars and black holes, but also during Type II supernovae 
    explosions that are the final stage in the evolution of massive stars. 
    The \astrobj{Sun} is an active star and cosmic ray flux variability is 
    a by-product of changes in the solar wind strength but solar wind 
    particles have much lower energies than the Galactic cosmic rays. On 
    the other hand, massive stars are short-lived ($<$ 20 Myr) and are 
    formed in great numbers during enhanced star formation episodes or 
    starbursts. An increase in the Galactic cosmic ray flux is therefore 
    expected within a few Myr of any starburst episode. The overall cosmic 
    ray flux on \astrobj{Earth} is affected not only by the neighbouring 
    emission rate, that is clearly proportional to the star formation rate, 
    but also by the magnetic field of the region where the \astrobj{Earth} 
    is located at a given time. This second effect correlates more with the 
    periodic crossing of regions of compressed gas than with star formation 
    itself and, in the absence of detailed modeling, might even dominate 
    over the neighbouring emission rate. This possibility is however going 
    to be neglected throughout the rest of the paper as compressed gas may 
    also be connected with star formation and because the effect of magnetic 
    fields on Galactic cosmic rays is weaker for higher energy particles (as 
    the ones expected from supernovae). Ney (1959) was the first to point 
    out that cosmic rays are the primary source for ionization in the 
    Troposphere. Recent empirical evidence confirms a direct connection 
    between average global low altitude cloud coverage and cosmic ray flux 
    (Svensmark and Friis-Christensen, 1997; Svensmark 1998, 2000; 
    Marsh and Svensmark, 2000; Palle Bago and Butler, 2000; 
    Wagner et al., 2001; Shaviv, 2001; Carslaw et al., 2002; Stozhkov, 2003). 
    An increase in the low altitude cloud coverage translates 
    into a decrease in the average global temperature and therefore 
    occurrence of ice age epochs. Christl et al. (2004) have found evidence 
    for a link between the flux of Galactic cosmic rays and climate during 
    the past 0.2 Myr. Svensmark (2004) has shown that variations in the flux 
    of Galactic cosmic rays at \astrobj{Earth} during the last 4.6 Gyr 
    indicate a remarkable resemblance to changes in the climate during the 
    same period of time suggesting again that \astrobj{Earth}'s climate 
    evolution is in some way linked to the evolution of the Galactic disk. 

    On the other hand, observations suggest that, throughout the 
    \astrobj{Milky Way}'s history, a non-negligible fraction of star 
    formation has occurred in starburst-like events (see Freeman and 
    Bland-Hawthorn, 2002, for a recent review). Shaviv (2003a) showed that 
    apparent peaks in the star formation rate of the 
    \astrobj{Solar Neighbourhood} coincide with ice age epochs on 
    \astrobj{Earth}. To support his claims, Shaviv uses results on the star
    formation rate in the \astrobj{Milky Way} disk obtained by Scalo (1987), 
    Barry (1988), and Rocha-Pinto et al. (2000a, b). 
    Although the majority of studies in this field use samples of stars
    in the \astrobj{Solar Neighbourhood} with no stars more distant than about
    100 pc being considered, it does not mean that the star formation
    history derived can only be applied to stars born in the 
    \astrobj{Solar Neighbourhood}. Nearby stars older than about 0.2 Gyr come 
    from birth sites which span a large range in Galactocentric distances.
    Wielen (1977) showed that the orbital diffusion coefficient 
    deduced from the observed increase of velocity dispersion with age
    implies that such stars have suffered an rms azimuthal drift of
    about 2 kpc for an age of 0.2 Gyr. Considerable, but smaller, drift
    should occur also in the radial direction. Wielen et al. (1996),
    on the basis of the \astrobj{Sun}'s metallicity and the radial 
    metallicity gradient in the Galactic disk, estimated that the 
    \astrobj{Sun} has migrated outward by 1.9$\pm$0.9 kpc in the past 
    4.5 Gyr. Shaviv also points out that 
    peaks in the star formation rate have to be connected with peaks in 
    the Galactic open cluster formation rate. Using data from the Lotkin 
    et al. (1994) open cluster catalogue he finds a main peak 300 Myr ago 
    and another, less statistically significant, 600 Myr ago. Mechanisms
    able to trigger enhanced star formation in the \astrobj{Milky Way} disk 
    suggested in his paper include gravitational tides induced during
    close encounters with the \astrobj{Magellanic Clouds} and interactions 
    with the Galactic spiral pattern. Shaviv (2003b) indicates that the
    actual picture could be even more complicated with variations in the 
    solar wind strength, greenhouse effect and cosmic ray flux modulation
    by a variable star formation rate in the \astrobj{Milky Way} disk 
    working together to model the long-term glacial activity on 
    \astrobj{Earth}. On the other hand, paleoclimatology records indicate 
    that the \astrobj{Earth}'s glaciation history 
    has shown a cyclic behavior during the last 1 Gyr with two additional 
    major glaciations 2.2-2.4 Gyr ago (mid-Proterozoic ice age) and 
    2.9-3.0 Gyr ago (Archean ice age). Paleoclimatological data from 
    Crowell (1999), Frakes et al. (1992) and Veizer et al. (2000) show 8 
    major ice age episodes during the last 1 Gyr. The timeline of the ice 
    age epochs from geological records is shown in Fig. \ref{iceage}.

    If, as paleoclimatological records suggest, the \astrobj{Earth}'s 
    glaciation history shows a cyclic behavior and this is connected with 
    the local star formation rate, an oscillating star formation history is
    expected. Early determinations of the star formation history of the
    \astrobj{Solar Neighbourhood} failed in finding a periodic star formation 
    rate. However, Hernandez et al. (2000) using stellar color-magnitude 
    diagrams found that the local star formation rate during the last 3 Gyr 
    has an oscillatory component of period $\sim$ 500 Myr superimposed on a 
    small level of constant star formation activity. This cyclic behavior is 
    interpreted by these authors as the result of repeated encounters with 
    the Galactic arm density pattern. Assuming that star clusters are the 
    elementary units of star formation (Clarke et al., 2000; Kroupa and Boily,
    2002; Lada and Lada, 2003), de la Fuente Marcos and de la 
    Fuente Marcos (2004) have derived a star formation history of the 
    Solar Circle (heliocentric distance $<$ 3.5 kpc) during the last 2 Gyr 
    that is fully consistent with the one derived by Hernandez et al. 
    (2000): cyclic star formation rate with a period of 0.4 $\pm$ 0.1 Gyr. 
    However, their results also suggest that for volume-limited open cluster 
    samples (the \astrobj{Solar Neighbourhood}, heliocentric distance 
    $<$ 1 kpc) the recovered history can be slightly different. The 
    \astrobj{Solar Neighbourhood} is defined as a volume centered on the 
    \astrobj{Sun} that is much smaller than the overall size of the 
    \astrobj{Milky Way} galaxy and yet large enough to contain a 
    statistically useful sample of stars (see, e.g., Binney and Tremaine, 
    1987). The appropriate size of the volume depends on which stars or 
    objects are going to be investigated: for white dwarfs, which are both 
    common and faint, it may consist of a sphere of radius 10 pc centred on 
    the \astrobj{Sun}, while for the bright but rare O and B stars, the 
    \astrobj{Solar Neighbourhood} may be considered to extend as far as 1 kpc 
    from the \astrobj{Sun}. Shaviv (2003a) suggests that a cyclic star 
    formation during the last 3 Gyr is in contradiction with the glaciation 
    records if the global scale climate changes on \astrobj{Earth} are 
    interpreted as a result of the increase in cosmic ray flux induced by 
    enhanced star formation. In this paper, we reanalyze this conclusion using 
    a volume-limited open cluster sample from the {\it Open Cluster Database} 
    (Mermilliod, 2004) corrected from evolutionary effects (open cluster 
    dissolution) as described in de la Fuente Marcos and de la Fuente Marcos 
    (2004). Our objective is to show that if a complete, volume-limited open 
    cluster sample is considered, the correlation, if real, has to be 
    recovered.
     
    This paper is organized as follows: in Section 2, we explain our
    choice of open cluster sample as well as present its properties. 
    The raw histogram of open cluster number vs. age and the corrections
    are also included in this Section as well as the potential limitations 
    of our approach. In Section 3, we compare the timing of the ice
    ages from geological records with the local star formation history
    inferred in Section 2. The statistical significance of the results
    is analyzed in Section 4. Open questions and conclusions are summarized 
    in Section 5.

 \section{Our sample: from the age distribution to the star formation rate}
    If star clusters (open clusters and associations) are the elementary 
    units of star formation in the \astrobj{Milky Way} disk they can, in 
    principle, be used to derive the star formation history, recent and old. 
    Unfortunately, stellar associations evolve and dissolve in a time-scale 
    of $\sim$ 50 Myr (Brown, 2002), therefore they cannot be used to study 
    the star formation rate. On the other hand, open clusters are 
    comparatively long lived objects that may serve as excellent tracers 
    of the structure and evolution of the Galactic disk. Although the, 
    now outdated, Lyng\aa \ catalogue (1987) has been the classical 
    reference on open cluster data for many years, the {\it Open Cluster 
    Database} (WEBDA) developed and maintained by J.-C. Mermilliod (1996, 
    2004) includes all the data already covered in Lyng\aa's catalogue and 
    many more. The latest update of the Open Cluster Database (WEBDA, 
    February 2004) includes 1731 open clusters with ages for 616 objects 
    (36\%). In this database we have found 568 clusters with age $\leq$ 2 
    Gyr. We consider this age cutoff because paleoclimatological data 
    indicate that cyclic glaciations started about 1 Gyr ago. Out of this 
    sample, 127 objects are within 1 kpc from the Sun and have $|z| \leq$ 
    100 pc (open cluster density, $\rho$ = 202 cluster/kpc$^3$). Only 3 
    clusters in this subsample are older than 1 Gyr. Unfortunately, our 
    initial choice for the volume produces an age distribution dominated by 
    low number statistics. In order to increase the statistical significance
    of our results, we have relaxed slightly the inclusion criteria. We
    have considered a cylinder of radius 1.5 kpc and $|z| \leq$ 150 pc
    with 256 open clusters and $\rho$ = 121 cluster/kpc$^3$. In our final
    sample, only 7 clusters are older than 1 Gyr.
   
    We have decided to consider a relatively small volume in order to
    minimize the effects of orbital diffusion. Special care has to be taken 
    when working with object samples in the Galactic disk as the stars or 
    open clusters in the sample considered, if old enough, may have formed 
    far away from their current location. In a non-volume-limited sample,
    the star/open cluster sample birth sites are in fact distributed over 
    a larger range of distances because of orbital diffusion. When trying
    to derive the global star formation history of the Galactic disk this
    process has very desirable consequences as the objects studied can 
    provide an estimate of the global star formation rate and their 
    conclusions can be extrapolated to the entire \astrobj{Milky Way} disk.  
    However, if we are studying an explicitly local process, samples can be
    polluted by objects that were not local at the time of their birth.
    In our case, we are trying to either confirm or reject a possible 
    correlation between the local star formation rate and the relatively 
    recent global changes in the \astrobj{Earth} climate. If a star or an 
    open cluster was not close enough to the \astrobj{Solar System} at the 
    time of a glaciation event it is very unlikely that it would have had 
    any effect on the recorded ice age episode. It is therefore of the main 
    importance to depopulate our sample of objects not born in the 
    \astrobj{Solar Neighbourhood} during the last 1 Gyr. It is basically 
    impossible to achieve this goal with field star samples but for open 
    cluster samples it should be not so difficult. Very little is known 
    about orbital 
    diffusion of entire star clusters but it is likely that the process is 
    significantly slower than for stars and inversely proportional to the
    mass of the open cluster. Our strategy is to use objects with 
    heliocentric distance $<$ 1.5 kpc and height scale $<$ 300 pc (or $|z| <$ 
    150 pc) to minimize the effects of pollution from orbital diffusion.  
    In addition to the sample presented, we have considered another five
    subsamples with more restrictive spatial inclusion criterion. Results
    from the various samples are fully consistent within the error limits.

%
%
\begin{figure*}
        \centerline{\hbox{
        \psfig{figure=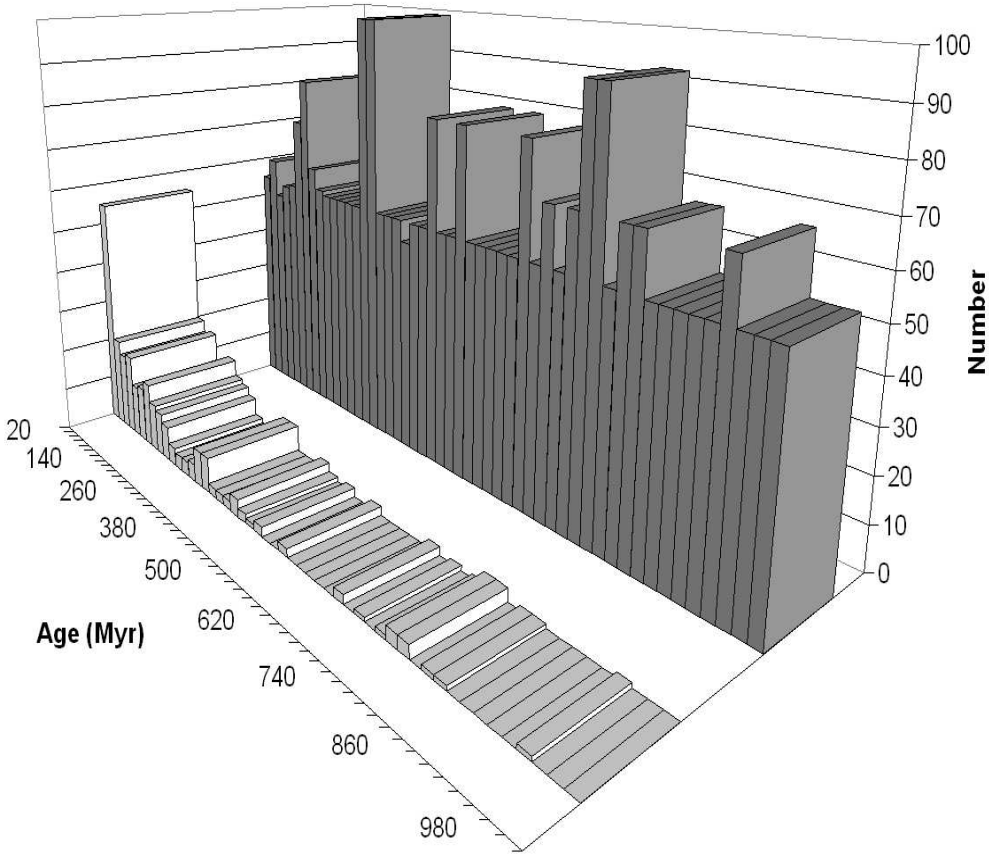,height=12cm,width=14cm,angle=0}
        }}
   \caption[]{Age distribution (white histogram) for a sample of 256 
              open clusters from the latest update (February 2004) of 
              WEBDA (Mermilliod, 2004) with age $\leq$ 1 Gyr, distance 
              from the Sun $\leq$ 1.5 kpc, and Galactic height $|z| \leq$ 
              150 pc. The gray histogram corresponds to the star 
              formation history of the \astrobj{Solar Neighbourhood} for
              an age bin of 20 Myr. This histogram has been corrected from
              evolutionary (cluster dissolution) effects. Volume and
              scale height corrections are not required as the open
              cluster sample is volume-limited and assumed complete.
             }
      \label{agedis}
\end{figure*}
%
%

    On the other hand, and as for any other physical parameter, open 
    cluster age determinations are affected by errors. In a large sample, 
    these errors are very likely to be non-homogeneous as different methods 
    have been used by different authors to calculate the ages. The disparity 
    in open cluster ages (when considering different authors, see WEBDA for 
    multiple examples) partly reflects the critical dependence of cluster 
    isochrone fitting on the adopted reddening (often large for open 
    clusters), even small reddening errors can create significant errors in 
    the derived age and metallicity. However, determining effective 
    temperatures and metallicities for cluster turn-off-stars directly 
    through echelle spectroscopy is free from systematic errors and subject 
    only to uncertainties in the model atmospheres. Some ages for clusters 
    in the sample considered have been determined using the first technique, 
    but others have been found using the second one or even other 
    indirect methods. It is relatively difficult, specially in papers older 
    than about 5 years, to find published estimations of the errors 
    associated with open cluster age determinations. The analysis in
    de la Fuente Marcos and de la Fuente Marcos (2004) shows that the average 
    age error is about 22\% and that younger ages are affected by larger 
    errors (although always $<$ 50\%). If this error sample can be considered
    as representative of the error range for the entire sample, the 
    youngest cluster (age $<$ 0.5 Gyr) errors are very likely in the range
    50-150 Myr with older cluster errors in the range 150-250 Myr.
    The age errors affect considerably the duration of the star formation 
    events, since they tend to scatter the ages of the star clusters 
    originally born in a burst. We can expect that this error could smear 
    out peaks and fill in gaps in the age distribution. Our conclusions are 
    mainly sensitive to the errors in age determination but also to the 
    degree of completeness of the open cluster sample. On the other hand, any 
    errors in theoretical stellar models also propagate into the results 
    obtained because the published open cluster ages always make reference to 
    theoretical stellar models, sometimes through direct isochrone comparison,
    others through the use of morphological features found in the cluster 
    color-magnitude diagram.

    Fig. \ref{agedis}, white histogram, is nothing more than the actual age 
    distribution of the open cluster sample considered in this paper, a
    raw histogram of open cluster numbers vs. age, of relatively young, 
    nearby open clusters. In order to interpret this distribution
    additional tools are needed. Following de la Fuente Marcos and
    de la Fuente Marcos (2004), we use results from Experimental Stellar 
    Dynamics to help us in this analysis by matching cluster dynamics and 
    cluster data to provide a better view of the recent star formation 
    history of the \astrobj{Solar Neighbourhood}. As described in
    de la Fuente Marcos and de la Fuente Marcos (2004) our numerical models
    have been calculated with the standard $N$-body code {\small NBODY5} 
    (Aarseth, 1985, 2003) for clusters located in the 
    \astrobj{Solar Neighbourhood}. These calculations include the effects 
    of stellar evolution, the Galactic tidal field, primordial binaries, 
    and realistic initial mass functions. These models indicate that an open 
    cluster has to include about 200-400 stars (at least) in order to survive 
    for about 0.5 Gyr, 400-700 to last 0.7 Gyr, 700-1000 to be detectable 
    after 0.9 Gyr and 1000-2000 to survive for about 1.3 Gyr. In Fig. 
    \ref{agedis}, we assume that the number of open clusters in each age bin 
    is, by hypothesis, correlated with the number of open clusters initially 
    born at that time. Once an age distribution is available, it is in 
    principle easy to recover the star formation history that gave rise to 
    the observed age distribution by using results from our realistic 
    $N$-body simulations. Our method does not assume any a priori galactic 
    structure or condition on the star formation rate and it basically 
    consists of three steps: (i) Construct a representative sample of open 
    clusters; (ii) Construct the age distribution diagram for the sample;
    (iii) Infer the star formation history from the diagram. As the life span 
    of larger clusters is longer, therefore, an usually high number of open 
    clusters at a given time interval can be interpreted as the result of an 
    event of enhanced star formation at that given age. If star cluster 
    masses are sampled from an open cluster initial mass function (hereafter 
    OCIMF), larger numbers translate into increased probability of formation 
    of large and therefore long-lived star clusters.  
        
    The resulting star formation history (Fig. \ref{agedis}, grey histogram)
    comes directly from the age distribution (Fig. \ref{agedis}, white 
    histogram), in an approach which assumes that the most frequent ages of 
    the open clusters indicate the epochs when the star formation was more 
    intense if we take into account that star clusters are the elementary 
    units of the star formation process. We have assumed that the open 
    cluster sample under study is only representative of the \astrobj{Solar
    Neighbourhood}, therefore the evolution of the local star formation rate 
    can be inferred from its age distribution, since the number of open 
    clusters in each age bin has to be correlated with the number of objects 
    initially at that time as a result of dissolution processes. As pointed 
    out before, the correction due to the dynamical evolution of the cluster 
    (cluster disintegration) is needed because our sample includes clusters 
    with different initial populations ($N_o$). The more massive clusters 
    have a longer life expectancy than the short-lived, small-$N$ clusters, 
    thus the latter would be missing in the older age bins. It is however 
    possible to correct for this effect using the OCIMF as well as results 
    on the open cluster lifetimes from numerical simulations. Fig. 
    \ref{iceage} shows the \astrobj{Solar Neighbourhood} sample with the 
    evolutionary correction described above (de la Fuente Marcos and de la 
    Fuente Marcos, 2004). Our method is only able to provide a lower limit 
    for the corrected number of objects per bin, therefore the actual number 
    is likely higher and the peaks, sharper. The effect is more important for 
    older ages.

 \section{Ice age epochs and local star formation rate}
    Paleoclimatological data indicate that in the time interval 1-2 Gyr
    no major ice age episodes have been recorded. Our star formation rate
    analysis suggests that during the same period, the local star formation
    rate was much lower than in the subsequent 1 Gyr: no open clusters
    appear in the time interval 1.2-1.9 Gyr. Within the error limits (about
    20\%), this is fully consistent with quiescent star formation (not
    enhanced) in the \astrobj{Solar Neighbourhood} during the age range 
    1-2 Gyr ago. However, and even before the Precambrian (600 Myr ago), 
    glaciation episodes have occurred at regularly spaced intervals of time 
    ($\sim$ 200 Myr) and lasting millions, or even tens of millions of years. 
 
%
\begin{figure*}
        \centerline{\hbox{
        \psfig{figure=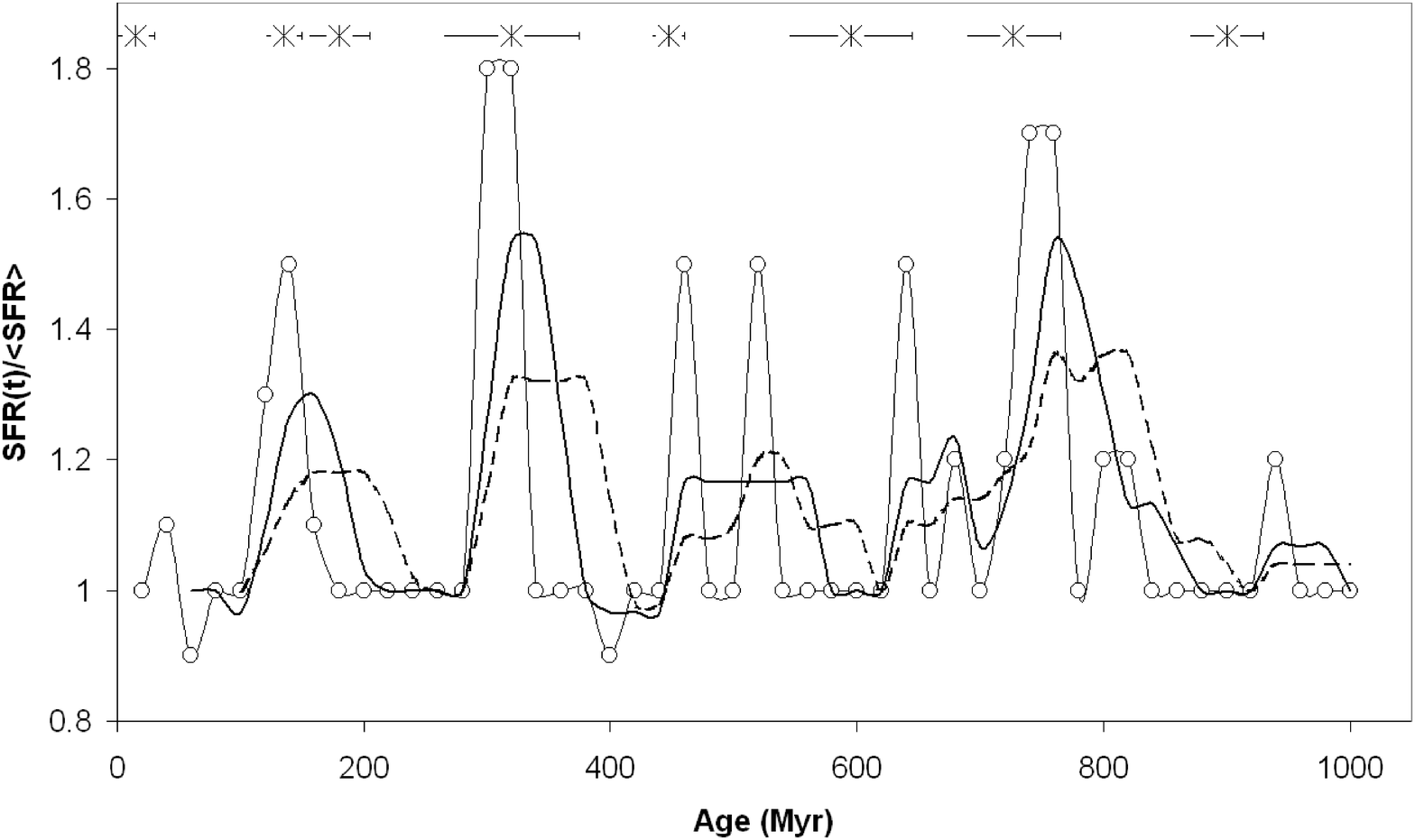,height=14cm,width=14cm,angle=0}
        }}
   \caption[]{Normalized star formation history of the Solar 
              Neighbourhood. The fitting lines are moving averages 
              with periods: 60 Myr (solid line) and 100 Myr 
              (dashed line). The position of glaciation epochs as 
              described by Crowell (1999) (Shaviv, 2003a, Fig. 10, 
              panel E) are marked. 
             }
      \label{iceage}
\end{figure*}
%
%
    Fig. \ref{iceage} suggests a strong correlation between enhanced star 
    formation episodes in the \astrobj{Solar Neighbourhood} and occurrence
    of ice ages on \astrobj{Earth}. Besides, it appears that glaciations 
    start some Myr after the (apparently) associated starburst began. 
    Nevertheless, the magnitude of the error in the ages of the open 
    clusters makes it difficult to confirm this apparent trend. On the 
    other hand, Shaviv (2003a) indicates that the Neoproterozoic era 
    (0.55-1 Gyr ago) was intrinsically cooler than the Phanerozoic (0-0.55 
    Gyr) era. If enhanced star formation is associated with glacial cycles,
    a higher average star formation rate is expected in the Neoproterozoic 
    period. This is exactly what is observed in Fig. \ref{iceage}. The 
    Marinoan and Varangian Glaciations (565, 615 Myr ago, respectively) and 
    the Sturtian Glaciation (725 Myr ago) have been the coldest epochs in 
    \astrobj{Earth}'s climate history with large, continental-size glaciers 
    covering enormous regions of the \astrobj{Earth}. Comparatively, the 
    mid-Mesozoic glaciation (160 Myr ago) was much less severe than the 
    others. On the other hand, paleoclimatological records suggest that in 
    the time interval 600-800 Myr the \astrobj{Earth} continuously had some 
    sort of ice age present. Geological evidence suggests that 750 Myr ago 
    glaciations reached sea level and equatorial latitudes ({\it Snowball 
    \astrobj{Earth}} scenario, Hoffman et al. 1995; Hoffman and Schrag, 2002). 
    If the strength of a given glaciation is a function of the strength of 
    the (assumed) associated enhanced star formation episode, then 
    Fig. \ref{iceage} is fully consistent with this scenario. On the other
    hand, Evans (2003) has found that Neoproterozoic glaciogenic sediments
    were deposited mainly at low paleolatitudes, in contrast to their 
    Pleistocene counterparts. High depositional latitudes dominate all
    Phanerozoic ice ages, exclusively low paleolatitudes characterize both
    of the major Precambrian glacial epochs. Transition between these
    two glaciation modes occurred within a 100 Myr interval, intrinsically
    coeval with the Neoproterozoic-Cambrian {\it explosion} of metazoan
    diversity. This study shows that glaciation episodes are much more 
    common from about 750 Myr ago than in the preceding sedimentary record.
    This evidence has been interpreted as a fundamental 
    Precambrian-Phanerozoic shift in \astrobj{Earth}'s glacial style.
   
 \section{Statistical significance of the results}
    Given the two sets of data, namely the epochs of recent enhanced star
    formation of the \astrobj{Solar Neighbourhood} and the glaciation 
    period record, Fig. \ref{iceage} is not enough to confirm or reject the 
    hypothesis of a causal correlation between the two phenomena. In this 
    section we will provide a much more detailed statistical comparison of 
    the two data sets.

    \subsection{Data Smoothing: Savitzky-Golay filters}
\begin{figure*}
        \centerline{\hbox{
        \psfig{figure=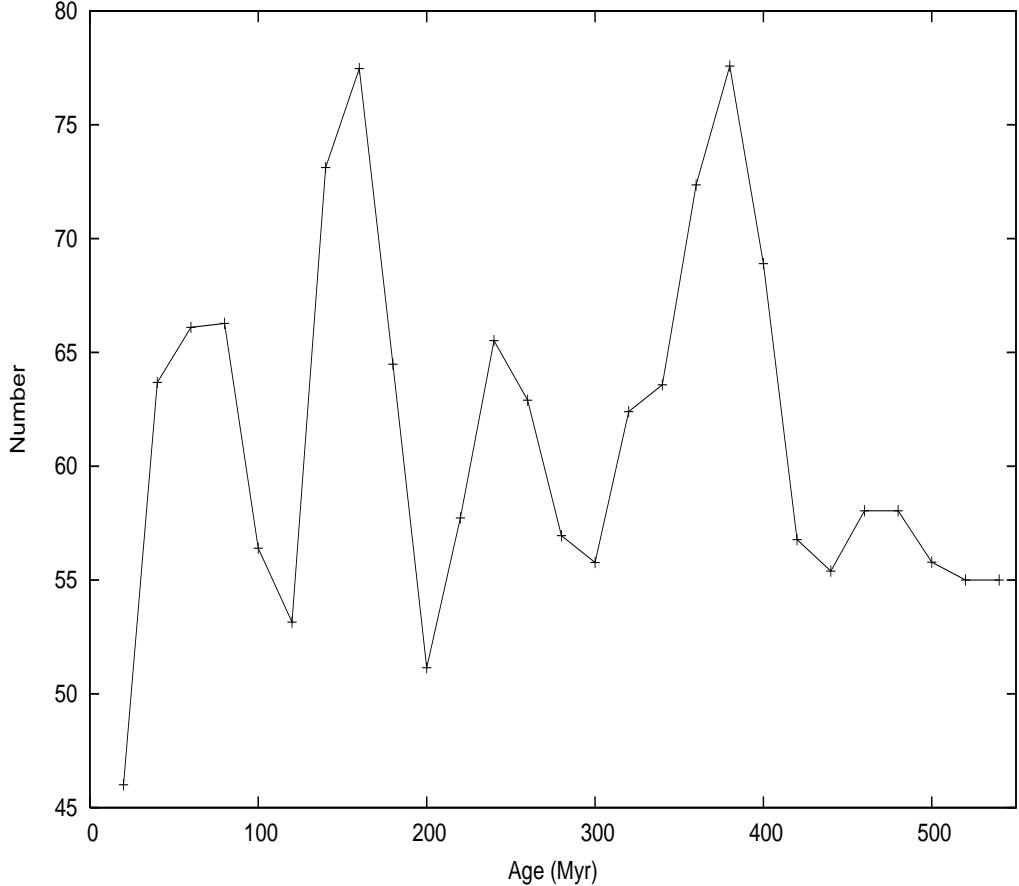,height=12cm,width=14cm,angle=-90}
        }}
   \caption[]{Result of smoothing the data by a Savitzky-Golay
              smoothing filter (of degree 2) using a window of
              9 points. While there is less smoothing of the
              broadest features, narrower features have their 
              heights and widths preserved.
             }
      \label{savgol}
\end{figure*}

\begin{figure*}
        \centerline{\hbox{
        \psfig{figure=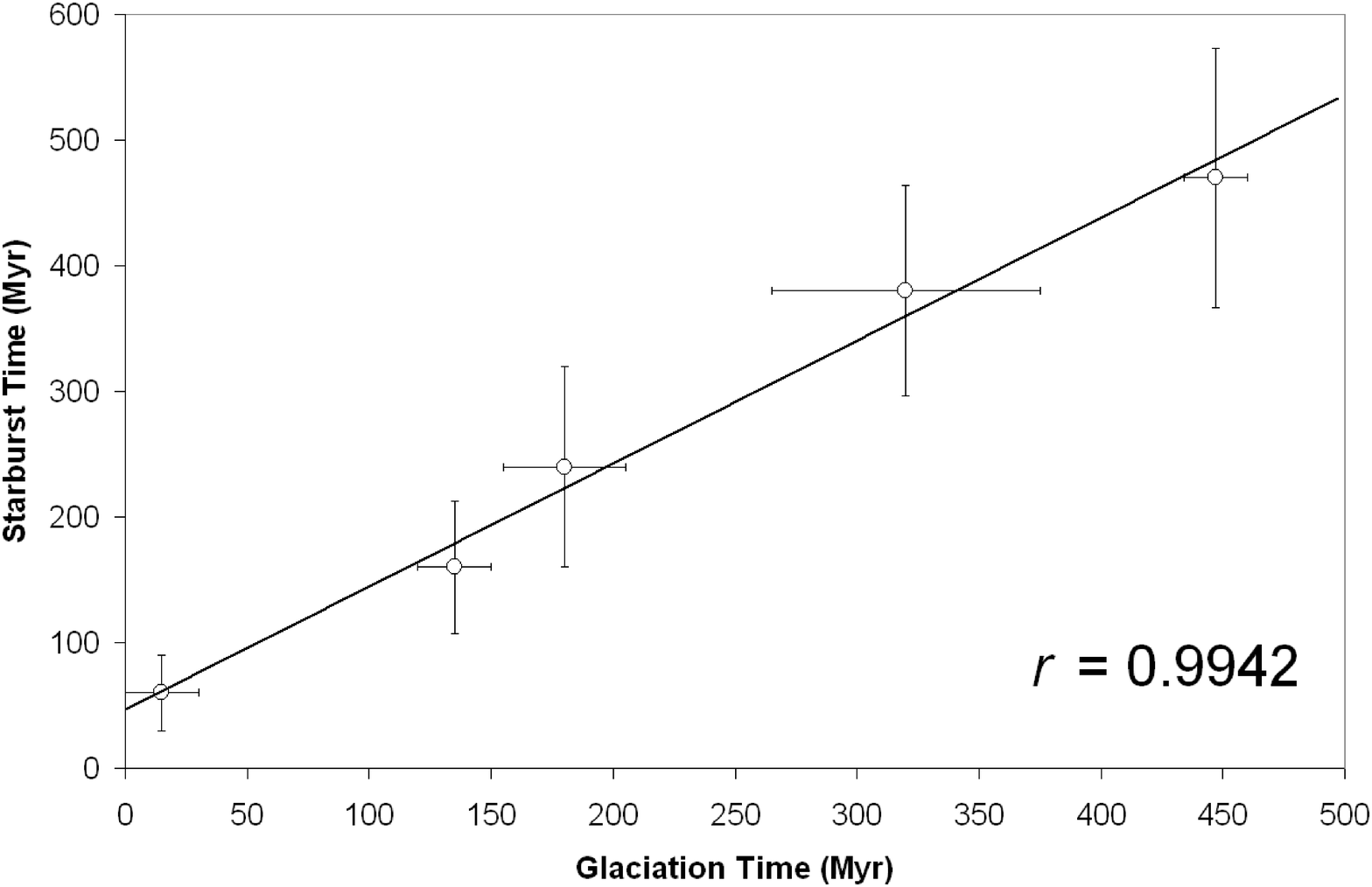,height=12cm,width=14cm,angle=0}
        }}
   \caption[]{Recovered enhanced star formation episodes as a function
              of the time-coincident glaciation events. The linear
              correlation coefficient, or Pearson's $r$, is 0.9942.
              A complete positive correlation is found within the error
              limits. The $x$-axis error bars correspond to the 
              characteristic size of the glaciation epoch, the $y$-axis
              error bars are consistent with the age errors discussed
              above. 
             }
      \label{sgcor}
\end{figure*}

       Data smoothing assumes that the variable under study is both slowly 
       varying and also corrupted by random noise. In our case, the star 
       formation rate is, in fact, changing relatively slowly with time and 
       the number of open clusters per age interval or bin is affected by 
       age determination errors and therefore corrupted. It is, however, 
       difficult to assume that age errors are a type of random noise because 
       important systematic errors may have been included during the age 
       determination process (de la Fuente Marcos and de la Fuente Marcos, 
       2004). In the previous sections we have used the moving window 
       averaging method to smooth the recovered star formation history. If 
       the underlying function is constant, or is changing linearly with time
       (increasing or decreasing), then no bias is introduced into the
       resulting smoothed curve. Higher points at one end of the averaging
       interval are on the average balanced by lower points at the other end.
       A bias is introduced, however, if the distribution function has a 
       nonzero second derivative: the height of local maxima are always 
       reduced and their width increased by moving window averaging. Narrow 
       features are broadened and suffer corresponding loss of amplitude.
       Savitzky-Golay filters (Savitzky and Golay, 1964), however, provide
       smoothing without loss of resolution. They approximate the underlying
       function by a polynomial (quadratic or quartic, typically) as 
       described by, e.g., Press et al. (1992). Fig. \ref{savgol} shows
       the recovered star formation history smoothed by a Savitzky-Golay
       filter (of degree 2) using 9 points. The filter recovers a missing 
       feature in the input distribution, the 200-300 peak.
       Five peaks of enhanced star formation are recovered 
       in the age range 0-550 Myr for the sample studied: 60$\pm$30 Myr, 
       160$\pm$53 Myr, 240$\pm$80 Myr, 380$\pm$84 Myr, and 470$\pm$103 Myr. 
       Here the errors have been estimated from the star cluster age errors. 
       The glaciation epochs in this age range are: 15$\pm$15 Myr, 
       135$\pm$15 Myr, 180$\pm$25 Myr, 320$\pm$55 Myr, and 447$\pm$13 Myr. 
       Both the paleoclimatological records of the glaciation episodes and 
       the timing of the enhanced star formation epochs appear to be fairly
       consistent. This can be interpreted as an indication of the existence 
       of a connection between the recent star formation record of the
       \astrobj{Solar Neighbourhood} and the glaciation period record. It may 
       be argued that Fig. \ref{sgcor}, although suggestive of a trend, 
       might not reflect the full nature of the correlation problem studied 
       due to overly optimistic errors for the cluster age determination 
       having being used. This is, however, not the case as the errors found 
       for the maxima are 50\%, 33\%, 33\%, 22\% and 22\%, respectively. 
       According to the age errors quoted in the literature (that can be 
       mainly found in papers published during the last five years) the upper 
       limit is about 50\% for young clusters ($<$ 100 Myr), 30\% for 
       intermediate age clusters (100-300 Myr), and 20\% for older clusters. 
       When considering these errors some of the starburst episodes overlap.
       The optimistic range would be about 10\%-20\% with the largest value 
       for the youngest objects. On the other hand, the linear correlation
       coefficient, or Pearson's $r$, is about 0.99 when we represent the
       maxima from the Savitzky-Golay smoothing procedure as a function of
       the time-coincident glaciation events, Fig. \ref{sgcor}. Our analysis 
       indicates the presence of a strong correlation between enhanced star 
       formation episodes and glaciation events during the last 550 Myr. 
%
%
%
%
%
%

    \subsection{Adaptative Smoothing: variable time kernel}
\begin{figure*}
        \centerline{\hbox{
        \psfig{figure=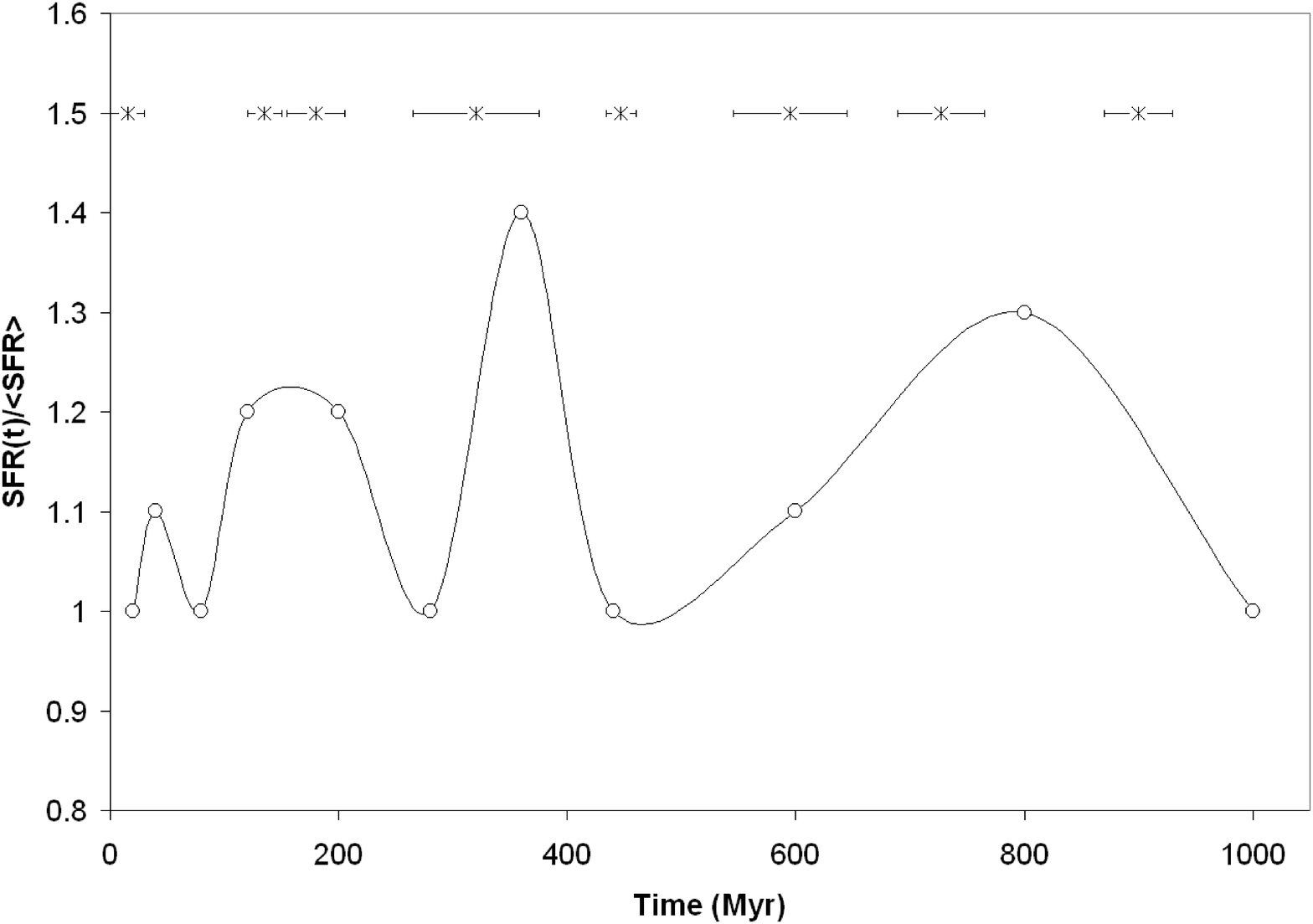,height=12cm,width=14cm,angle=0}
        }}
   \caption[]{Result of smoothing the data by a variable time kernel 
              smoothing filter. Although the overall curve is very
              smooth, narrower features have their heights and widths 
              altered. In particular, the amplitude of the oldest maximum
              is likely to be much higher. The position of glaciation 
              epochs as described by Crowell (1999) (Shaviv, 2003a, Fig. 10, 
              panel E) are marked. 
             }
      \label{moverr}
\end{figure*}

\begin{figure*}
        \centerline{\hbox{
        \psfig{figure=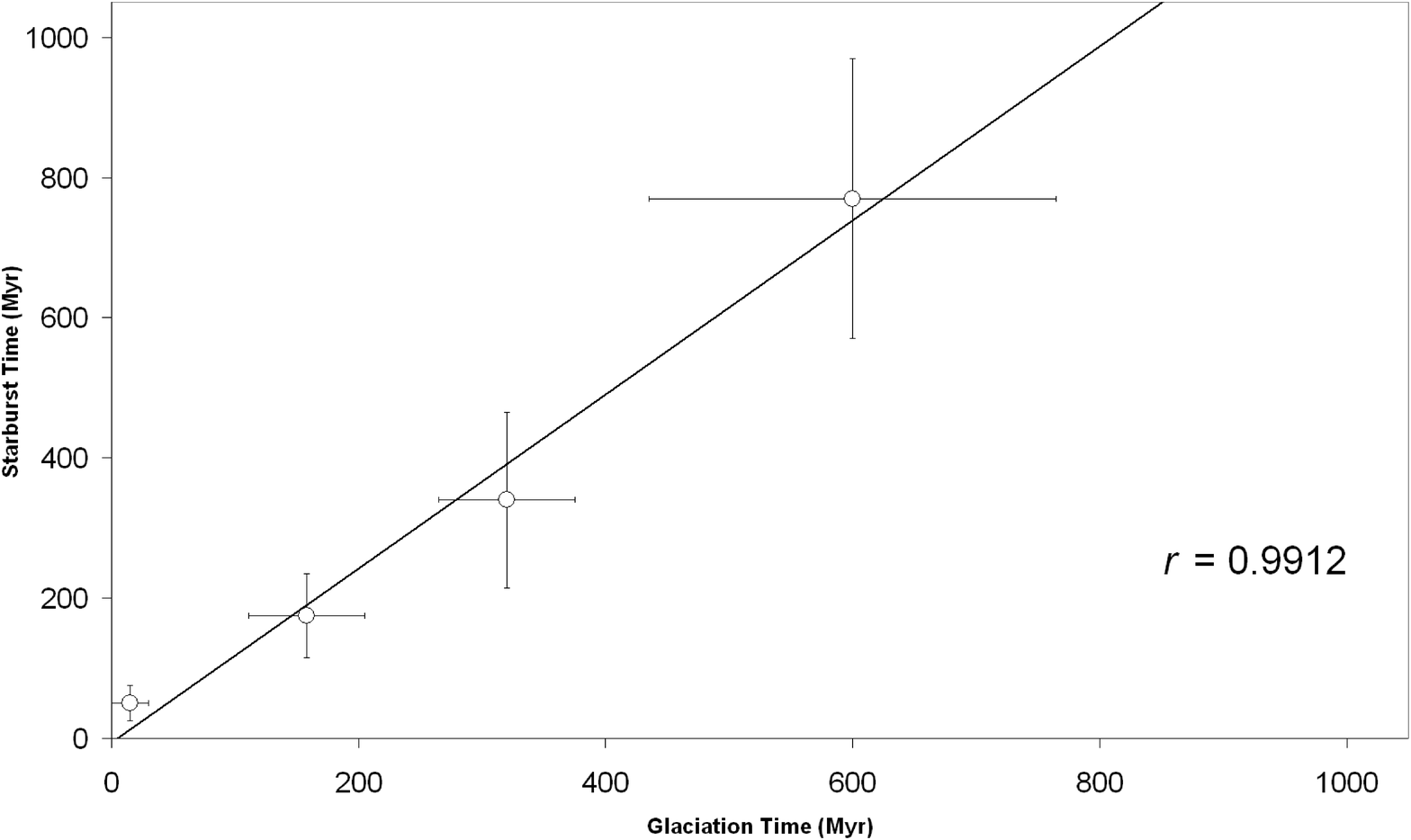,height=12cm,width=14cm,angle=0}
        }}
   \caption[]{Recovered enhanced star formation episodes as a function
              of the time-coincident glaciation events. The linear
              correlation coefficient, or Pearson's $r$, is 0.9912.
              A complete positive correlation is found within the error
              limits. The $x$-axis error bars correspond to the (degraded)
              characteristic size of the glaciation epoch, the $y$-axis
              error bars are consistent with the age errors discussed
              above. 
             }
      \label{lincor}
\end{figure*}

       In Figs. \ref{iceage} and \ref{savgol}, both the moving average and 
       the Savitzky-Golay filter applied to the results do not directly 
       reflect the errors in cluster ages. These errors could be as high as 
       50\% for young clusters (age $<$ 100 Myr), around 30\% for
       open clusters in the age range 100-600, and $\sim$20\% for the
       age range 600-1000. The errors have been, however, included in the
       determination of the peaks of enhanced star formation activity. In
       this section, a variable time kernel for the smoothing following the 
       actual errors has been applied to the recovered data. Fig. 
       \ref{moverr} shows the result of our {\it adaptative smoothing} as 
       well as the glaciation epochs. The obtained star formation history 
       can be considered as the worst case scenario with the most pessimistic 
       evaluation for the errors being considered and included in the 
       smoothing process. Fig. \ref{moverr} indicates that the uncertainties 
       in open cluster age determination for ages older than about 500 Myr 
       are clearly too large at present, in principle, to draw any conclusion 
       regarding either the presence or the absence of a correlation
       with the glaciation events. At this stage it is not possible to
       resolve this problem, but by degrading the resolution of the 
       paleoclimatological data we should be able to obtain not a proof but
       at least an indication of the presence of a correlation. 
%
%
%
       We have done this to obtain Fig. \ref{lincor}. 
       The peaks have then been linearly correlated to the glaciation 
       record for the four remaining maxima in the star formation 
       history. Fig. \ref{lincor} shows the obtained correlation that
       is fairly good within the error limits. Although, due to the
       errors in cluster ages, only four events can be correlated, the
       result suggests a strong causal connection between the recent star 
       formation history of the \astrobj{Solar Neighbourhood} and the 
       glaciation period record. The choice of the glaciation events
       included in Fig. \ref{lincor} has been fully constrained by the
       errors considered in cluster age determinations.
%
%
%
 
    \subsection{Model optimization: Saha's $W$ merit function} 
       When attempting to answer the question of whether or not two data
       sets are drawn from the same distribution function, the chi-square
       test is the natural choice for binned distributions. However, if 
       the number of points per bin for one of the distributions is small,
       the implicit assumption in the chi-square test (the data object
       points follow a Gaussian distribution) is no longer valid. This is
       exactly what we have in our open cluster age histogram with most
       of the bins in the age range 500-1000 Myr containing 0, 1, 2 or
       3 clusters. Saha (1998) developed a test to study this special case.
       Saha's test is based on the merit function:
       \begin{equation}
          W = \prod_{i = 1}^{B} \frac{(m_i + s_i)!}{m_i!\ s_i!} \,,
       \end{equation}
       where the $i$th bin has $m_i$ model points and $s_i$ data object
       points. The above expression can be used even when shot noise
       is not negligible. The idea behind the test is the following, for
       a given data object set and binning, we can change the parameters
       of the model to generate a $m_i$ model points distribution to  
       maximize $W$. In our case, the data object set is the open cluster
       sample and the model to be adjusted is the glaciation period record.
       If the model glaciation period record that maximizes $W$ resembles
       the paleoclimatological record, then the correlation can be considered
       positive. In the following, we have used the test for the age
       interval 300-1000 Myr. Unfortunately, this is the time range where
       errors are in the range 100-220 Myr. We have constructed three merit
       functions for a bin size of 50, 100 and 200 Myr, respectively. The
       number of clusters observed in a particular bin follows the
       distribution in Fig. \ref{agedis}, white histogram. The model 
       following the glaciation period record has been constructed as 
       follows: i) we assume that enhanced star formation episodes are
       characterized by the formation of $N_c$ clusters per unit time, 
       ii) these clusters have life times longer than the time interval 
       studied (in this work, 1 Gyr), iii) the number of clusters to be
       observed in a given bin is proportional to the fraction of
       glaciation episode included in that bin as well as to the duration
       of the episode, as in Fig. \ref{iceage}. The total number of model 
       clusters generated under these assumptions is about 20\% larger than 
       the actual number of observed clusters to simulate incompleteness of 
       the observed sample. For the first merit function (bin size, 50 Myr), 
       we have performed 1,000,000 Monte Carlo simulations of random 
       glaciation records against the open cluster age record. The 
       {\it random glaciation records} include the same number of events 
       that the one matching the observed glaciation record. Only 4\% of
       the simulated sets produce a better model-data value of the merit
       function conserving the number of glaciation episodes in the
       studied age range. The glaciation intervals that maximize the merit
       function are: 300-425, 500-550, 600-650, 700-775, and 800-900 in Myr. 
       The actual glaciation record in the age range considered is: 300-375, 
       434-460, 545-645, 689-765, and 870-930 in Myr. A similar result is
       obtained for the larger bin sizes. 

 \section{Discussion and conclusions} \label{dis}
    In this paper we have attempted to confirm or reject the existence
    of a correlation between the local star formation history as derived
    from the open cluster formation rate in the \astrobj{Solar Neighbourhood} 
    and the timing of glaciation episodes on \astrobj{Earth}. Peaks in the 
    age distribution diagram of the cluster sample were interpreted as 
    signatures of starbursts. Our results indicate that the analyzed star 
    formation rate presents two components: periodic episodes 
    of enhanced star formation superimposed on a quiescent star formation 
    level. Although it could have been induced by tidal interactions with the
    Magellanic Clouds (Shaviv, 2003a; de la Fuente Marcos and de la Fuente 
    Marcos, 2004), there is, however, a cyclic behavior in the
    burst sequence that may be better explained by the density wave
    hypothesis (Lin and Shu, 1964) for the presence of spiral arms in
    late-type galaxies. A model like the one outlined in 
    Hernandez et al. (2000a) and developed in Martos et al. (2004) can 
    explain easily the 0.4 Gyr periodicity found by de la Fuente Marcos
    and de la Fuente Marcos (2004) for the Galactic disk.
    For a pattern speed $\Omega_p$ = 20 km s$^{-1}$ kpc$^{-1}$
    (e.g. Martos et al., 2004) and a Galactocentric distance 
    $R_{\odot}$ = 8.5 kpc it implies an orbital period of about 1 Gyr
    for the \astrobj{Sun}. If the enhanced star formation episodes are, in 
    fact, due to the interactions with the spiral arms it means that our 
    \astrobj{Galaxy} has two arms. This has been recently suggested by 
    Martos et al. (2004) using 
    a completely different approach. However, the periodicity observed for 
    the local star formation rate is 
    150$^{+100}_{-50}$ and it could be 
    interpreted as evidence in favour of the presence of a four-armed 
    spiral pattern in the Galactic disk. Unfortunately, this conclusion is 
    strongly 
    affected by the significant value of the associated errors that make
    it also compatible with a two-armed spiral pattern. The different
    periodicity observed for the \astrobj{Solar Neighbourhood} can also
    be interpreted as higher harmonics due to local substructures.

    For our complete, volume-limited open cluster sample in the \astrobj{Solar
    Neighbourhood} only 7 objects are found in the age range 1-2 Gyr,
    with no objects in the age range 1.2-1.9 Gyr. However, for an open
    cluster sample in the Solar Circle (heliocentric distance $<$ 3.5 kpc)
    the recovered star formation rate is rather different, with several
    peaks of star formation. This apparently unexpected result brings
    to our attention the fact that in order to trigger enhanced star
    formation two ingredients are required: a mechanism able to trigger
    the process and a certain number of giant molecular clouds available
    to host the process. The distribution of giant molecular clouds in the
    Milky Way disk is not uniform and therefore this may translate into 
    non-uniform star formation when considering relatively small regions
    ($<$ 1.5 kpc) and young ($<$ 2 Gyr) ages. The discrete behavior should
    disappear when both longer time scales and larger volumes are
    considered. On the other hand, the star formation histories recovered 
    from the two descriptions do not need to be consistent as the effects 
    of both spiral pattern and galactic interactions (two plausible 
    mechanisms able to trigger enhanced star formation) are not uniform 
    through the entire disk. In this work we have only used a fairly 
    small part of the disk trying to minimize the effects of orbital 
    diffusion. It implies a restriction of the studied region not only in 
    volume but also in time to avoid pollution by objects formed far away 
    from the 
    studied volume. This can be considered a {\it Lagrangian approach} 
    to the star formation history of the Milky Way disk.

    It is important to point out that the timing coincidences and the 
    overall good correlation found between the recent star formation
    rate in the \astrobj{Solar Neighbourhood} and the glaciation period 
    record on \astrobj{Earth} are totally independent of the assumed 
    model for the spiral
    structure (two, four-armed) and the spiral pattern speed. Our analysis
    not only suggests a strong correlation in the timing of the events
    (enhanced star formation and glaciation episodes), but also in the
    severity and length of the episodes. On the other hand, the correlation
    between the meteoritic data (exposure ages of Fe/Ni meteors) given
    in Fig. 10, panel F (Shaviv, 2003a) and the normalized local star 
    formation rate shown in Fig. \ref{iceage} is fairly good, with the 
    lowest levels in meteor exposure indicating lower cosmic ray flux.

    Although the exact causes for glaciation episodes, and the ice ages 
    cycles within them, are still far from being well established, the
    astrophysical data discussed in this paper strongly suggest that
    galactic dynamics plays a primary role in the long-term evolution of
    global climate on \astrobj{Earth}. The overall evolution is however 
    likely the result of a complicated dynamic interaction between the 
    Solar Cycle, periodic changes in the orbit of the \astrobj{Earth}, 
    natural oscillations of the ocean/continent/atmosphere system, 
    variations in the composition of the atmosphere, and Galactic dynamics. 
    Even though our conclusions are uncertain as they are based on a
    relatively small subsample of a larger but still incomplete 
    sample composed of almost 2000 open clusters, they are similar to the 
    ones previously found by other independent studies. 

 \begin{ack}
   The authors are particularly grateful to Sverre Aarseth for his helpful
   comments and suggestions. In preparation of this paper we made use of the 
   Open Cluster Database, operated at the {\it Institut d'Astronomie de 
   l'Universit\'e de Lausanne}, Switzerland, the NASA Astrophysics Data 
   System and the ASTRO-PH e-print server. An anonymous referee made a 
   number of extremely helpful suggestions which improved the paper 
   significantly.
 \end{ack}

\end{document}